\documentclass[preprint,nofootinbib,pra]{revtex4}

\usepackage{graphicx}
\usepackage{amsmath}
\usepackage{dcolumn}
\usepackage{subcaption}
\usepackage{comment}
\usepackage[section]{placeins}
\usepackage{titlesec}
\usepackage{graphicx} 
\usepackage{amsmath} 
\usepackage{amssymb} 

\usepackage{natbib} 
\usepackage{hyperref}

\begin{document}

\title {Thermodynamics of effective loop quantum black holes}

\author{F. G. Menezes$^1$, H. A. Borges$^1$, I. P. R. Baranov$^2$, S. Carneiro$^{1,3}$}

\affiliation{$^1$Instituto de F\'{\i}sica, Universidade Federal da Bahia, 40210-340, Salvador, BA, Brazil\\$^2$Instituto Federal de Educa\c c\~ao, Ci\^encia e Tecnologia da Bahia, 40301-015, Salvador, BA, Brazil\\$^3$Observat\'orio Nacional, 20921-400, Rio de Janeiro, RJ, Brazil}

\date{\today}

\begin{abstract}

We study the thermodynamics of a non-singular black hole model with effective quantum corrections motivated by Loop Quantum Gravity (LQG). The effective geometry has a transition surface that connects trapped and anti-trapped regions with the same mass. There is a minimum mass for which the horizon temperature and Komar energy are zero, and the black hole stops its Hawking evaporation. For horizons above this limit, we present the grey-body factors, emission spectra, and the mass loss rate, solving a one-dimensional Schrödinger-type equation with an effective short-range potential barrier for massless fields of spins $0$, $1/2$, $1$ and $2$.

\end{abstract}

\maketitle

\section{Introduction}

Shortly after discovering the thermodynamic laws of black holes, Stephen Hawking demonstrated that, by analysing a scalar field near the event horizon, black holes can emit particles in a process now known as Hawking radiation \cite{Wald1999}. This emission follows a law similar to black body radiation, where the temperature of the black hole is proportional to the surface gravity at the horizon. The key difference between black hole emission and black body radiation is the presence of grey-body factors. These factors account for the probability of particles overcoming the potential barrier and escaping to infinity.

In the context of General Relativity, Hawking radiation is a continuous process. The heat capacity of a Schwarzschild black hole is negative and inversely proportional to its mass. This means that as the black hole loses mass through Hawking radiation, its temperature increases, which accelerates the rate of evaporation. Ultimately, this process leads to the complete disappearance of the black hole.

Quantum field theory (QFT) describes the strong and weak forces and electromagnetism at both high and low energies, while general relativity (GR) describes gravity. A key difference between these theories is their treatment of spacetime. Quantum theory uses a fixed spacetime background, whereas GR describes spacetime as dynamic and shaped by matter and energy. This difference makes QFT and GR fundamentally incompatible. Developing a quantum theory of gravity to unify them has been a focus in recent decades. Loop Quantum Gravity (LQG) is one promising approach, offering a non-perturbative, background-independent framework \cite{lewandowski,livros1,livros2,livros3}.

Effective theories outline key features of underlying quantum theories. Effective models of black holes with quantum corrections inspired by LQG offer precise descriptions, but some of them face limitations, such as incomplete extensions between internal and external regions, no asymptotic flatness, quantum effects not limited to high curvature areas, and the loss of covariance \cite{PRL,AOS,mariam,bojowald}.

To address these issues, a quantization scheme for the spherically symmetric vacuum solution was recently proposed, which employs a particular polymerisation procedure \cite{espanhois,bascos2}. It results in a covariant, hyperbolic, non-singular spacetime, where the classical singularity is replaced by a transition surface between black and white holes of equal masses. The transition surface radius can be fixed by making use of the LQG area gap, imposing a minimal area condition \cite{Fernando}. In this way, the model can be extended to Planck-scale black holes, expanding it beyond previous effective models limited to macroscopic black holes.

In a previous work of some of the authors \cite{Fernando2}, we have shown that remnant solutions are possible in the context of this single polymerisation parameter model (ABBV model). In the present paper we investigate their thermodynamic properties, as their heat capacity, and also calculate the grey-body factors, which allow computing the black hole mass evolution.

The paper is structured as follows. In the next section we present the uni-parametric polymerised model and the solutions matching the minimal area condition. In Section~III the Komar energy and the horizon temperature are re-derived, leading to the black hole heat capacity. Section~IV is devoted to the computation of the grey-body factors and the emission rate of Hawking radiation, which allows us to obtain the black hole mass evolution. In Section~V we present our conclusions.

\section{The effective model \label{secII}}

In the canonical variables used in \cite{PRL,AOS}, the classical Hamiltonian \cite{bascos3,rakesh,esteban} can be written as
\begin{equation}\label{eq:hcl}
    H_{\rm cl}=-\frac{1}{2 G \gamma}\left[\left(b+\frac{\gamma^2}{b} \right) p_b + 2c p_c    \right],
\end{equation}
corresponding to the interior homogeneous classical metric
\begin{equation} \label{homogeneous}
    ds^2 = - N^2 dT^2 + \frac{p_b^2}{p_c} dx^2 + p_c d\Omega^2,
\end{equation}
with the lapse function given by $N = \gamma \sqrt{p_c}/b$.
From it we construct the effective Hamiltonian through the polymerisation \cite{florencia}
\begin{equation}\label{eq:polymerisation}
b \rightarrow \dfrac{\sin\left(\delta_b b\right)}{\delta_b}, \quad \quad \quad p_b \rightarrow \dfrac{p_b}{\cos(\delta_b b)},
\end{equation}
and by including the regularisation factor
\begin{equation}\label{eq:regularisation}
\dfrac{\cos(\delta_b b)}{\sqrt{1+\gamma^2\delta_b^2}},
\end{equation}
which leads to
\begin{align}\label{eq:abbvhamiltonian}
H_{\rm eff}
&= -\dfrac{1}{2G\gamma\sqrt{1+\gamma^2\delta_b^2}}\left[  \left(\dfrac{\sin(\delta_b b)}{\delta_b} +\frac{\gamma^2 \delta_b}{\sin{(\delta_b b)}}\right)p_b + 2cp_c\cos(\delta_b b)\right].
\end{align}

It can be shown that $M$, defined as
\begin{equation}\label{eq:b}
    \frac{\sin^2(\delta_b b)}{\gamma^2 \delta_b^2}=\frac{2M}{\sqrt{p_c}}-1,
\end{equation}
is an integration constant of the Hamilton equations. It is also possible to show that the exterior static metric, which is asymptotically flat, is given by \cite{espanhois,bascos2}
\begin{equation} \label{metric}
   ds^2 = - \bigg(1-\frac{2M}{r}\bigg) dt^2 + \bigg(1-\frac{2M}{r}\bigg)^{-1}\bigg(1-\frac{r_0}{r}\bigg)^{-1} dr^2 + r^2 d\Omega^2,
\end{equation}
where $d\Omega^2=d\theta^2+\sin^2\theta d\phi^2$ is the metric on the unitary $2$-sphere in polar
coordinates, $M$ can be identified with the mass of the black hole, and $r_0$ with a minimum radius. Notice that when $r\gg r_0$ the metric reduces to the Schwarzschild geometry.

In the effective quantum spacetime, the classical singularity is replaced by a transition surface interpolating black and white hole interiors of the same mass, with horizons at $r_h = 2M$. 
The transition surface is solution of the equation $\dot{p}_c=0$, with radius
\begin{equation} \label{r0}
    r_0 = \sqrt{p_{c}^{\rm min}}=2M\frac{\gamma^2\delta_b^2}{b_0^2},
\end{equation}
with
\begin{equation} \label{Qmax}
b_0=\sqrt{1+\gamma^2\delta_b^2},
\end{equation}
where $\gamma$ is the Barbero-Immirzi parameter.
In the classical limit, the polymerisation parameter $\delta_b$ goes to zero and the transition surface vanishes, recovering the classical singularity.

Now we can obtain a specific dependence of $\delta_b$ on the horizon radius by fixing the area of the transition surface to be equal to the LQG area gap \cite{modesto}, so that $r_0 =\sqrt{\sqrt{3}\gamma}$ is independent of $M$. Then, from ($\ref{r0}$) we find the simple relation
\begin{equation}
\label{DeltabEquals1}
    \gamma^2\delta_b^2=\frac{1}{r_h/r_0-1},
\end{equation}
with real polymerisation parameter for $r_h \geq r_0$.

\section{The black hole thermodynamics}
\label{secIII}

\begin{figure}[t]
\centering
\includegraphics[width=10cm]{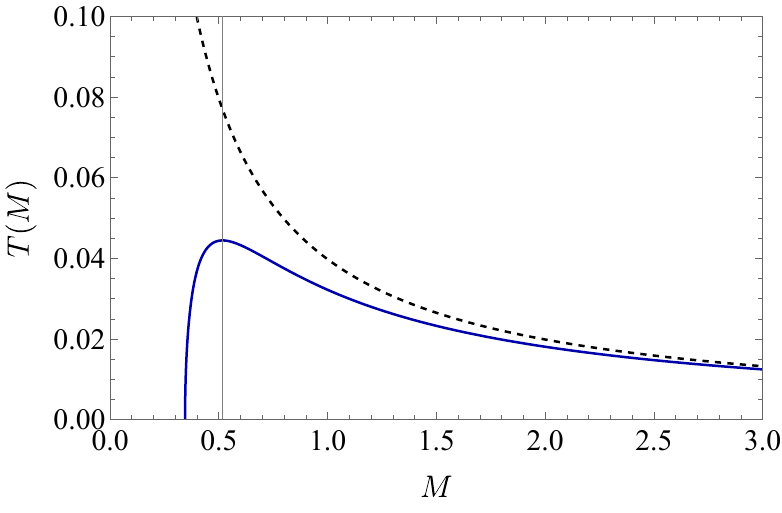}
\caption{\label{fig:bhtemperature} Black hole temperature as a function of mass for this effective model (solid line) and for the classical solution (dashed line).}
\end{figure}

Since the metric components in \eqref{metric} are time-independent, there are time-like Killing vectors associated with stationary observers, such that the surface gravity is given by \cite{Fernando2}
\begin{equation}
\label{eq:sphericallySG}
    \kappa  = \frac{1}{2r_h}\sqrt{1-\frac{r_0}{r_h}}.
\end{equation}
Due to Hawking radiation, the surface gravity can be related to the temperature 
of the black hole as
\begin{equation}\label{Te}
T =\frac{\kappa}{2\pi}. 
\end{equation}
Notice that, as shown in Fig.~1, the temperature reaches a maximum when $r_h=3r_0/2$, and it vanishes at the minimum mass $M_{\text{min}} = r_0/2$, when $r_h \rightarrow r_0$.

A proper definition of the energy enclosed by a spherical surface of radius $r_h$ is given by the Komar energy at the horizon \cite{Komar1959}. It is straightforward to obtain \cite{Fernando2}
\begin{equation}
\label{komarenergy}
    E_K(r)=\frac{r_h}{2}\sqrt{1-\frac{r_0}{r_h}},
\end{equation}
which is proportional to the surface gravity. This energy equals the mass of the black hole $M$ in the Schwarzschild limit, while for the quantum corrected metric \eqref{metric} it is smaller and goes to zero when the horizon approaches the minimum radius.

To compute the heat capacity of the black hole we evaluate the Komar energy derivative with respect to the temperature, obtaining
\begin{equation}
    C=-\frac{A}{2}\bigg(\frac{2r_h - r_0}{2r_h - 3r_0}\bigg).
\end{equation}
As shown in Fig.~2, the heat capacity is negative if the horizon radius is larger than $3r_0/2$. 
For $r_h \gg r_0$ it reaches the well-known result $C=-A/2$, that is, if the black hole absorbs energy its temperature decreases. At the critical value $2r_h = 3r_0$, corresponding to the maximum temperature, there is a phase transition and the heat capacity diverges. Beyond this value it changes sign and becomes positive as the temperature decreases with the evaporation of the black hole, reaching $A/2$ when $r_h \rightarrow r_0$.

\begin{figure}[t]
\centering
\includegraphics[width=10cm]{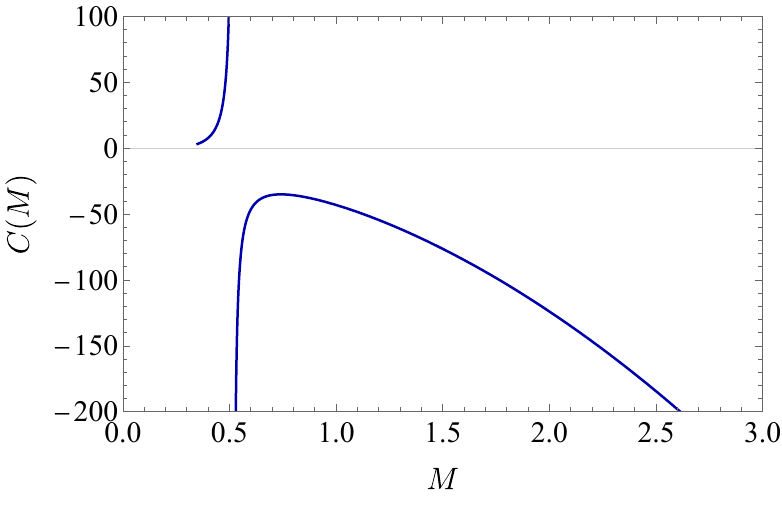}
\caption{\label{fig:HeatCapacity} {The heat capacity as a function of the black hole mass, showing a discontinuity at the point of maximum temperature $M = 3r_0/4$.}}
\end{figure}

\section{GREY-BODY FACTORS AND EMISSION RATES}

We know that a black hole can emit thermal radiation with a blackbody spectrum generated at the horizon due to quantum fluctuations \cite{Wald1999}. As it propagates, the geometry surrounding the black hole modifies the shape of the spectrum for an observer located at spatial infinity. Basically, the geometry of spacetime acts as a potential barrier that filters Hawking radiation in such a way that part of it will tunnel through the potential and the rest will be reflected back into the hole. To obtain the quantity that measures how much the spectrum deviates from a pure blackbody spectrum, called the grey-body factor, defined in terms of the transmission and reflection coefficients, one needs to solve the equations of motion for matter fields with the boundary conditions.

To see how we can obtain the potential, consider the propagation of test quantum fields in the presence of the black hole effective metric ($\ref{metric}$). For a massless scalar field, the simplest case, the dynamics is described by the Klein-Gordon equation 
\begin{eqnarray}
\label{eq:kleing}
\nabla_\mu\nabla^\mu \Phi = \frac{1}{\sqrt{-g}}\partial_\mu \left( \sqrt{-g}g^{\mu\nu}\partial_\nu \Phi\right) = 0,
\end{eqnarray}
where, as usual, $g$ denotes the determinant of the metric.

Since the metric has spherical and time symmetries, it is convenient to expand the scalar field as
\begin{eqnarray}
\label{eq:fieldecomposition}
\Phi(t,r,\theta,\varphi) = \frac{1}{r}\sum_{l,m}\psi_{l\omega}(r)Y_{lm}(\theta,\varphi)e^{-i\omega t},
\end{eqnarray}
where $Y_{lm}(\theta,\phi)$ are the spherical harmonics. Then, the radial equation can be transformed into a Schr\"odinger wave equation
\begin{equation}\label{sch}
\bigg[\frac{d^2}{dr^2_{*}}+w^2-V_s(r)\bigg]\psi_s=0,
\end{equation}
where $\omega$ is the energy of the field and $V_s$ is the short-range potential for spin $s=0$ given by
\begin{equation}\label{p1}
V_0=\bigg(1-\frac{2M}{r}\bigg)\bigg[\frac{l(l+1)}{r^2}+\frac{4M+r_0}{2r^3}-\frac{3Mr_0}{r^4}\bigg].
\end{equation}
The tortoise coordinate for metric ($\ref{metric}$) is defined as
\begin{equation}
dr_{*}=\bigg[\sqrt{1-\frac{r_0}{r}}\bigg(1-\frac{2M}{r}\bigg)\bigg]^{-1}dr,
\end{equation} 
such that the black hole horizon is located at $r_{*}=-\infty$, and $r_{*}=\infty$ when $r=\infty$. 

\begin{figure}[t]
\begin{subfigure}{.5\textwidth}
  \centering
\includegraphics[width=.98\linewidth]{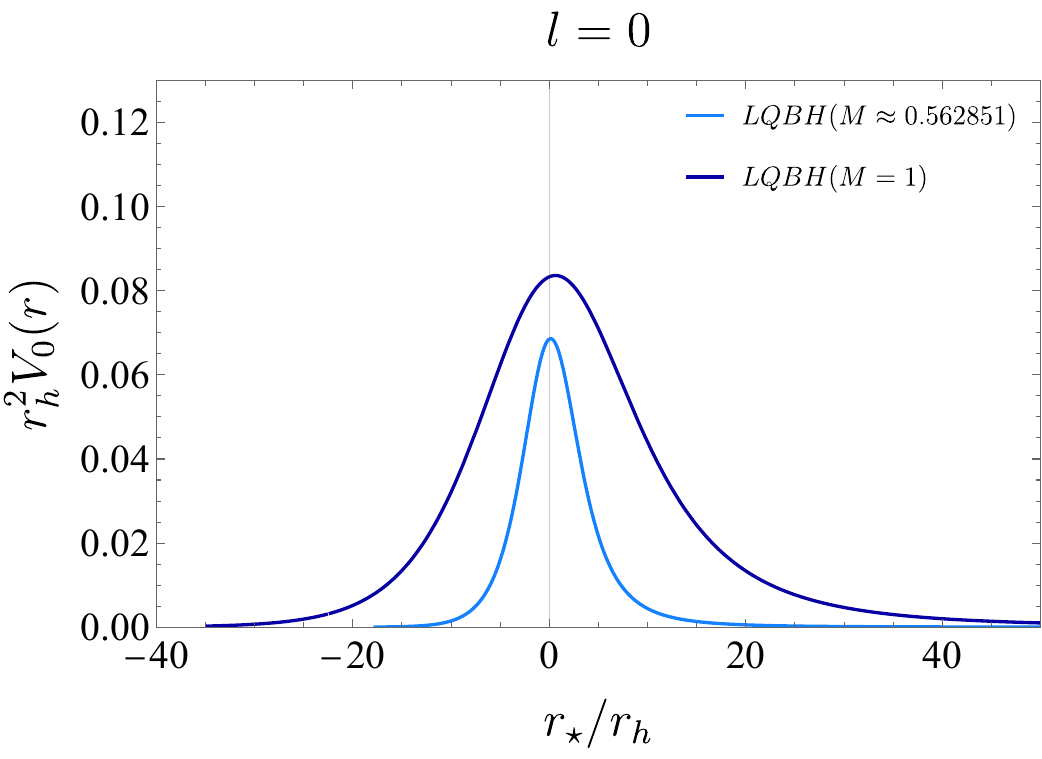}
\end{subfigure}%
\begin{subfigure}{.5\textwidth}
  \centering
\includegraphics[width=.98\linewidth]{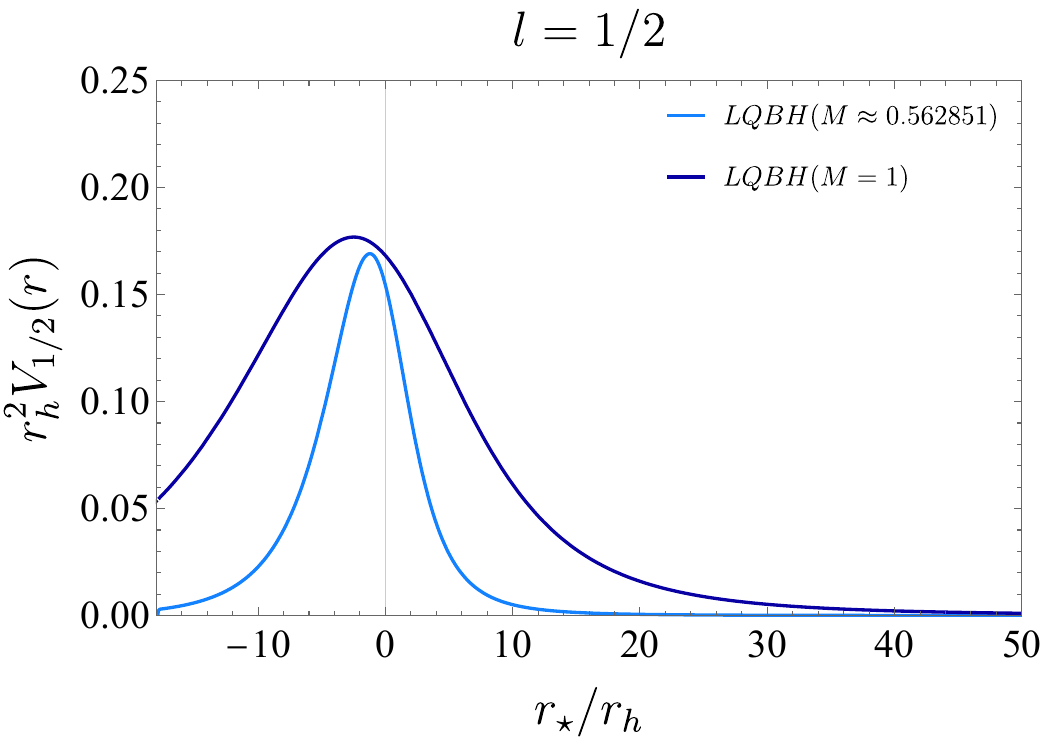}
\end{subfigure}\\

\begin{subfigure}{.5\textwidth}
  \centering \includegraphics[width=.98\linewidth]{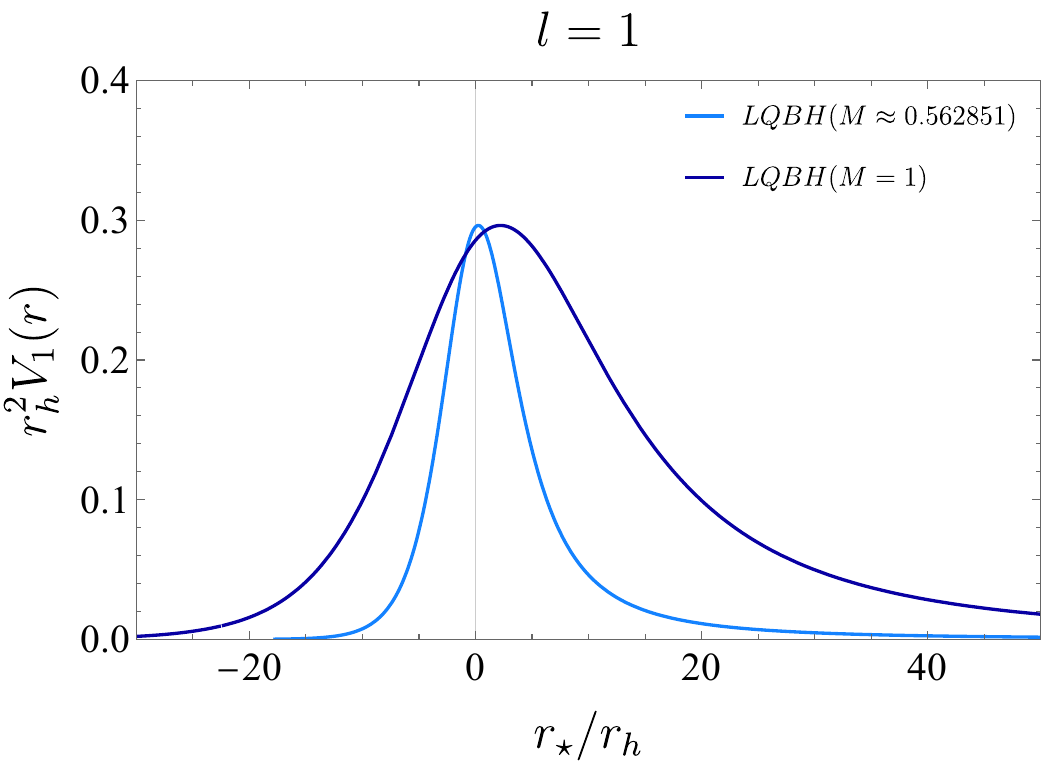}
\end{subfigure}%
\begin{subfigure}{.5\textwidth}
  \centering
\includegraphics[width=.98\linewidth]{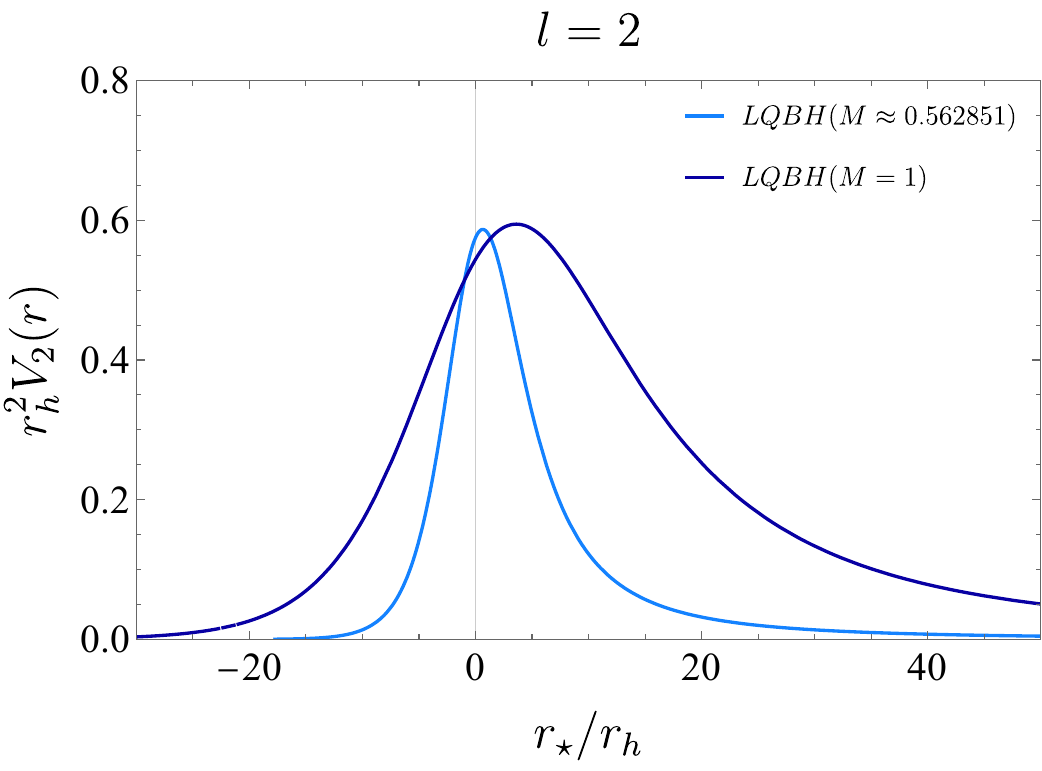}
\end{subfigure}
\caption{The effective potentials for the first mode of each spin field.}
\label{fig:Potentials}
\end{figure}

For higher spin fields, the procedure for obtaining a Schr\"odinger-type wave equation is not so straightforward. 
Instead, a reformulation of the equations of motion using a null tetrad field, known as the Newman–Penrose formalism, results in the Teukolsky master equation, valid for all spins. After suitable transformations, the Teukolsky radial equation can be transformed into the Schr\"odinger equation (\ref{sch}) \cite{Blackhawk2}, with potentials for spin fields $1$, $2$ and $1/2$ given respectively by
\begin{equation}\label{p2}
V_1=\frac{l(l+1)}{r^2}\bigg(1-\frac{2M}{r}\bigg),
\end{equation}
\begin{equation}\label{p3}
V_2=\bigg(1-\frac{2M}{r}\bigg)\bigg[\frac{l(l+1)}{r^2}-\frac{(12M-5r_0)}{2r^3}+\frac{7Mr_0}{r^4}\bigg],
\end{equation}
\begin{equation}\label{p4}
V_{1/2}=\bigg(1-\frac{2M}{r}\bigg)\bigg[\frac{l(l+1)+1/4}{r^2}\pm\sqrt{l(l+1)+1/4}\sqrt{1-\frac{r_0}{r}}\bigg(\frac{M}{r^3\sqrt{1-2M/r}}-\frac{1}{r^2}\sqrt{1-\frac{2M}{r}}\bigg)\bigg],
\end{equation}
where $l=s, s+1, ...$ are the waves angular modes.

In Fig. \ref{fig:Potentials}, we show these potentials for masses $M=3r_0/4$, corresponding to the maximum temperature, and $M=1$ for the first mode $l=s$. Note that for spin $1/2$ there are two solutions. We choose the solution of positive sign in equation (\ref{p4}) since the use of the other solution does not alter the main results of this work.

\begin{figure}[t]
\begin{subfigure}{.5\textwidth}
  \includegraphics[width=.98\linewidth]{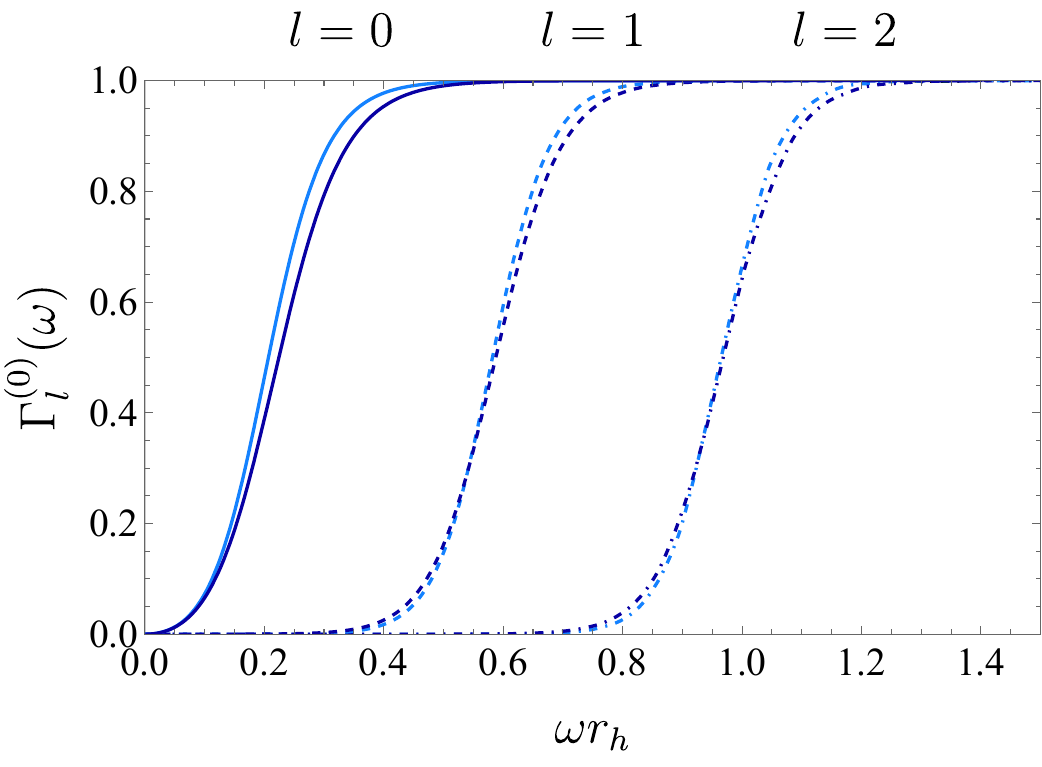}
\end{subfigure}%
\begin{subfigure}{.5\textwidth}
  \includegraphics[width=.98\linewidth]{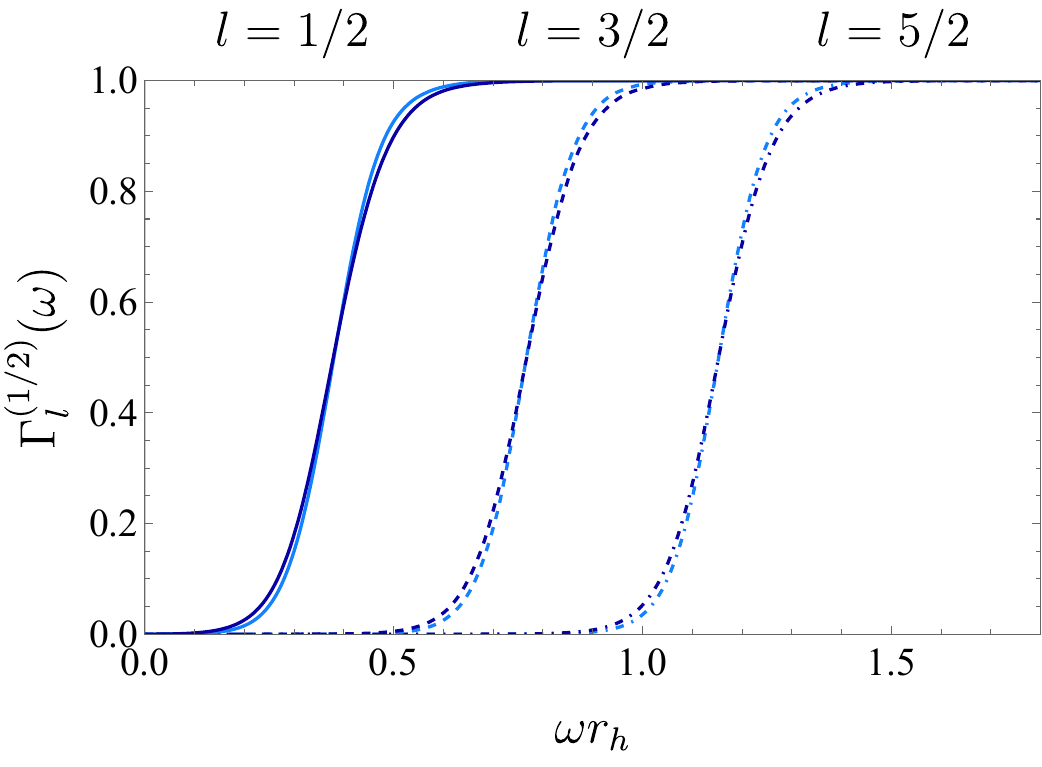}
\end{subfigure}\\
\begin{subfigure}{.5\textwidth}
  \includegraphics[width=.98\linewidth]{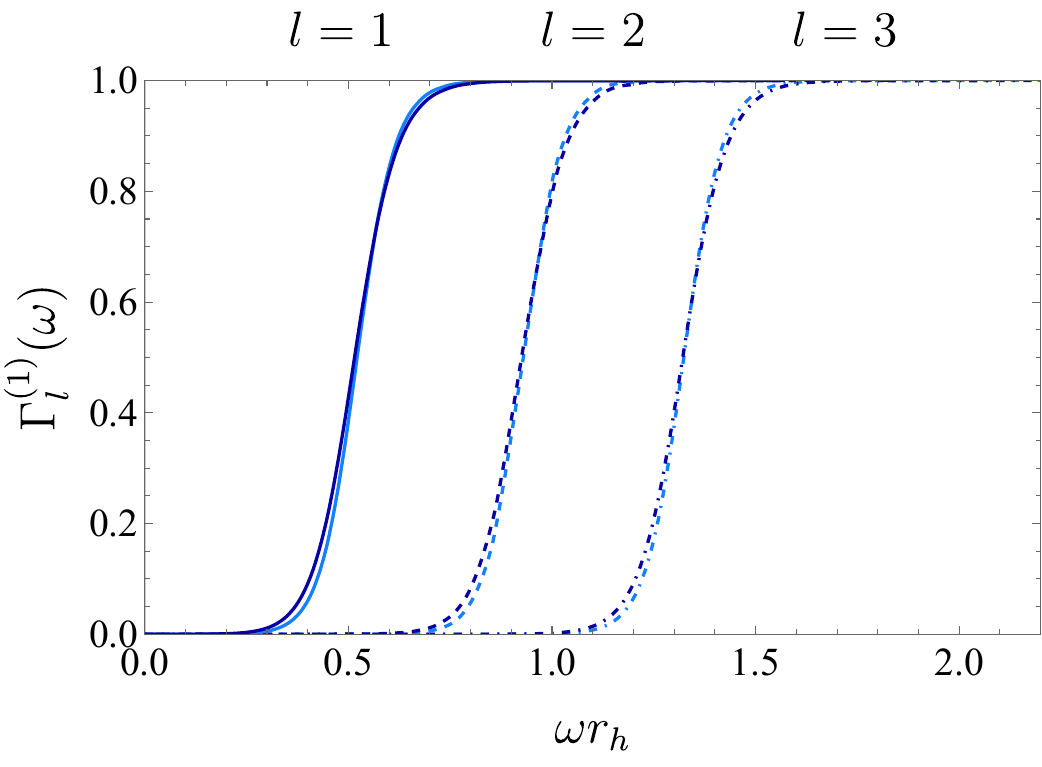}
\end{subfigure}%
\begin{subfigure}{.5\textwidth}
  \includegraphics[width=.98\linewidth]{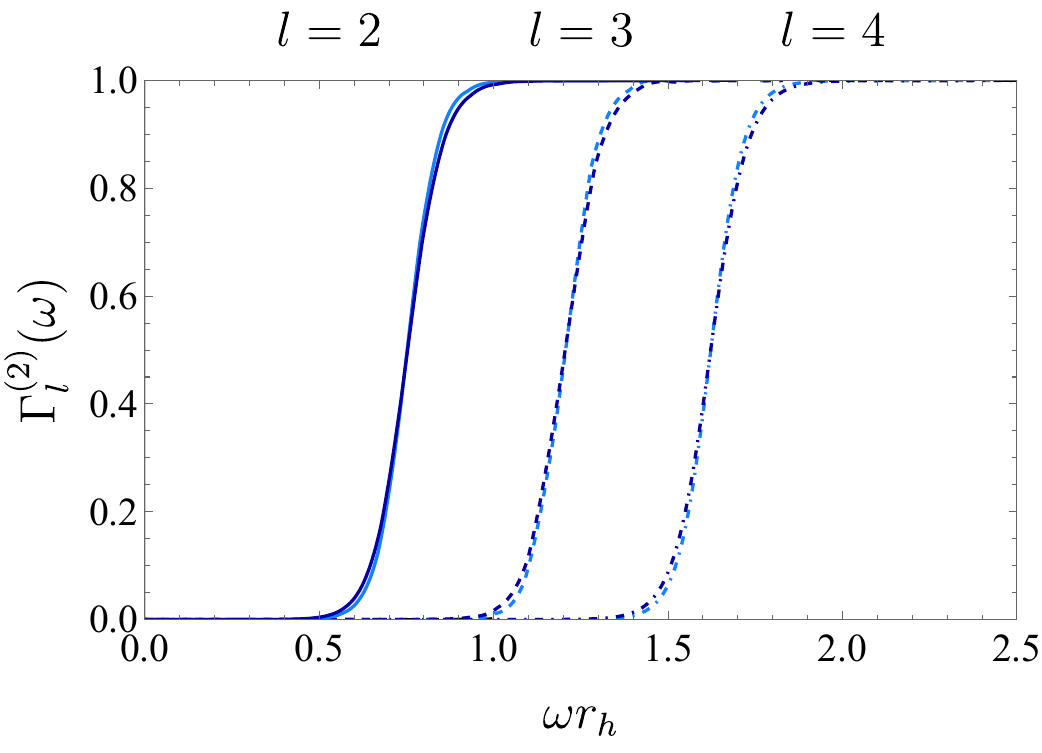}
\end{subfigure}
\caption{ Grey-body factors $\Gamma^{s}_l$ for spins $s=0$, $1/2$, $1$, $2$ as functions of $\omega\, r_h$ for the first three modes (solid, dashed and dot-dashed lines). Masses $M=0.56$ and $M=1$ (light and dark blue lines, respectively) are used for the effective black hole.}
\label{fig:GF}
\end{figure}

If we consider asymptotic solutions of the wave equation for an incoming wave originating at infinity, that satisfies the boundary conditions
\begin{equation}
\psi\sim e^{i\omega r_{*}}+Re^{-i\omega r_{*}}, \ \ \ \ r_{*}\rightarrow +\infty,
\end{equation}
\begin{equation}
\psi\sim Te^{-i\omega r_{*}}, \ \ \ \ r_{*}\rightarrow -\infty,
\end{equation}
where $R$ and $T$ are the reflection and transmission coefficients, respectively, then the grey-body factor is defined as
\begin{equation}
\Gamma^{s}_{l}(M, \omega)=\left|\frac{T}{R}\right|^2.
\end{equation}
After computing the grey-body factors, we can obtain the number of particle species $i$ of spin $s$ characterized by $g_i$ degrees of freedom emitted per unit time and energy,
\begin{equation}
\frac{d^2N_i^s}{dtd\omega}=\frac{g_i}{2\pi}\sum_l\frac{\Gamma^{s}_{l}}{e^{\omega/T}-(-1)^{2s}}.
\end{equation}
Because each particle carries energy $\omega$, the total energy emitted per unit time is
\begin{equation}
\frac{dE}{dt}=\int_0^{\infty}\sum_s\frac{d^2N_i^s}{dtd\omega}\omega d\omega.
\end{equation}

\begin{figure}[t]
\begin{subfigure}{.5\textwidth}
  \centering
  \includegraphics[width=1\linewidth]{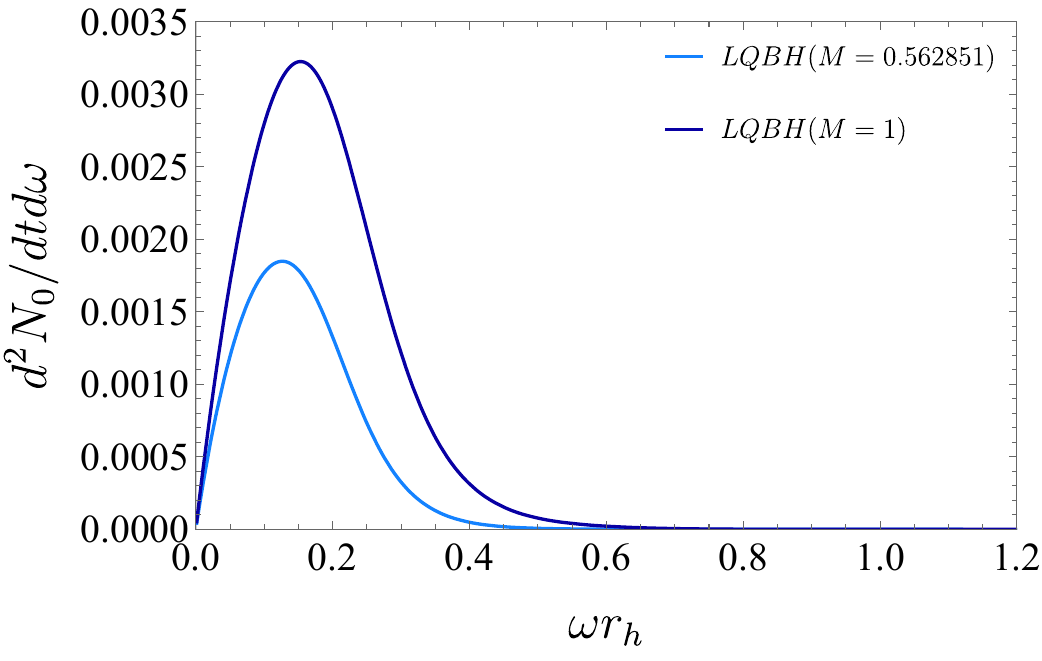}
\end{subfigure}%
\begin{subfigure}{.5\textwidth}
  \centering
  \includegraphics[width=1\linewidth]{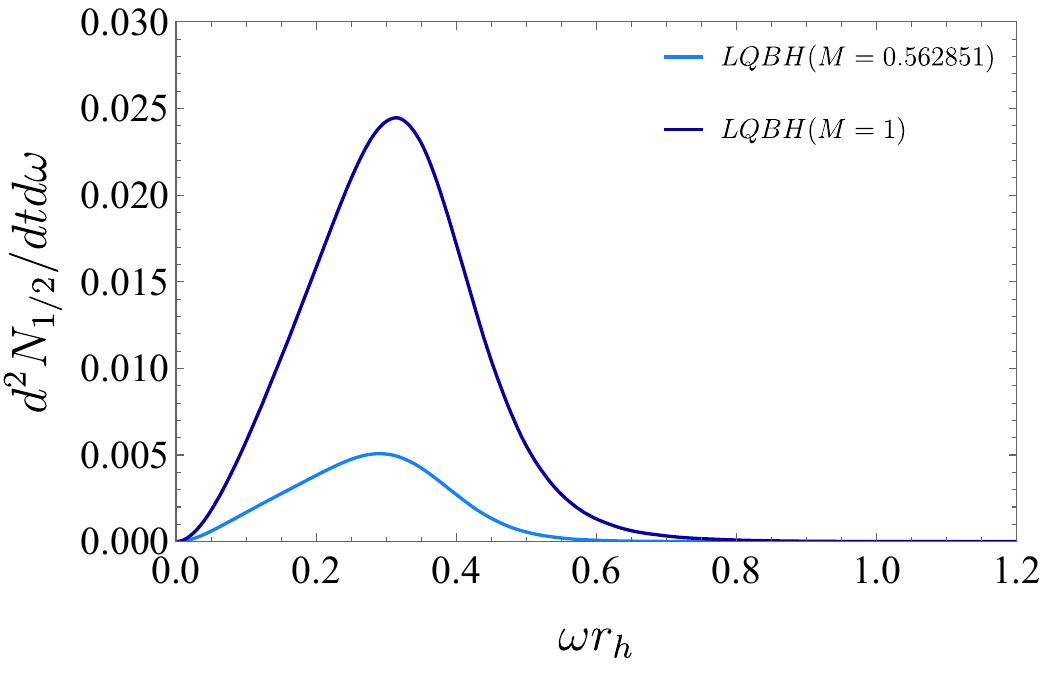}
\end{subfigure}\\
\begin{subfigure}{.5\textwidth}
  \centering
  \includegraphics[width=1\linewidth]{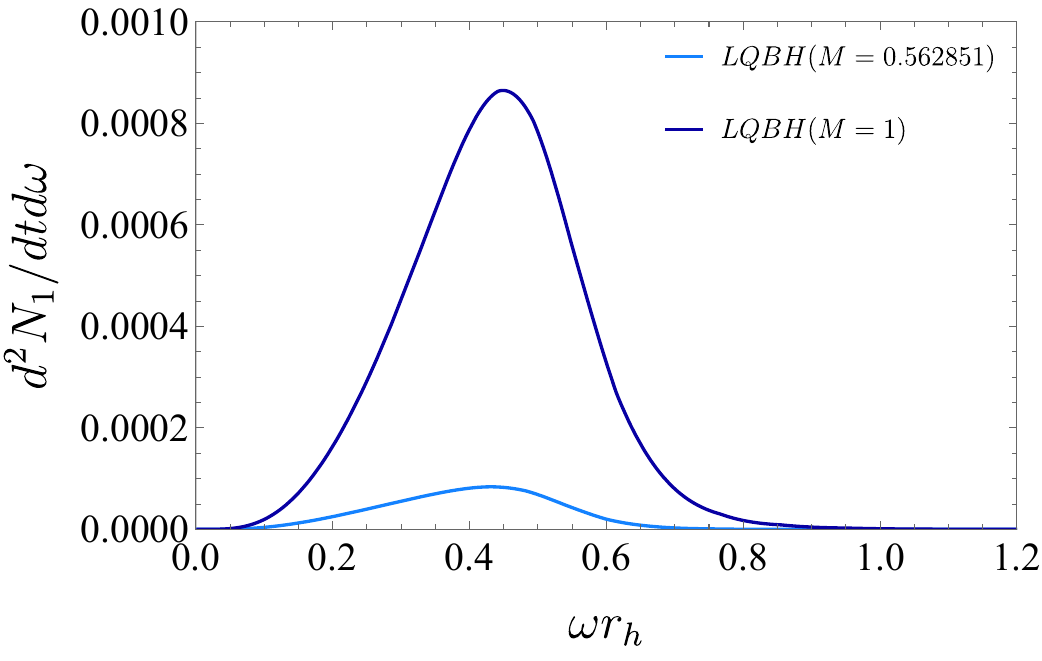}
\end{subfigure}%
\begin{subfigure}{.5\textwidth}
  \centering
  \includegraphics[width=1\linewidth]{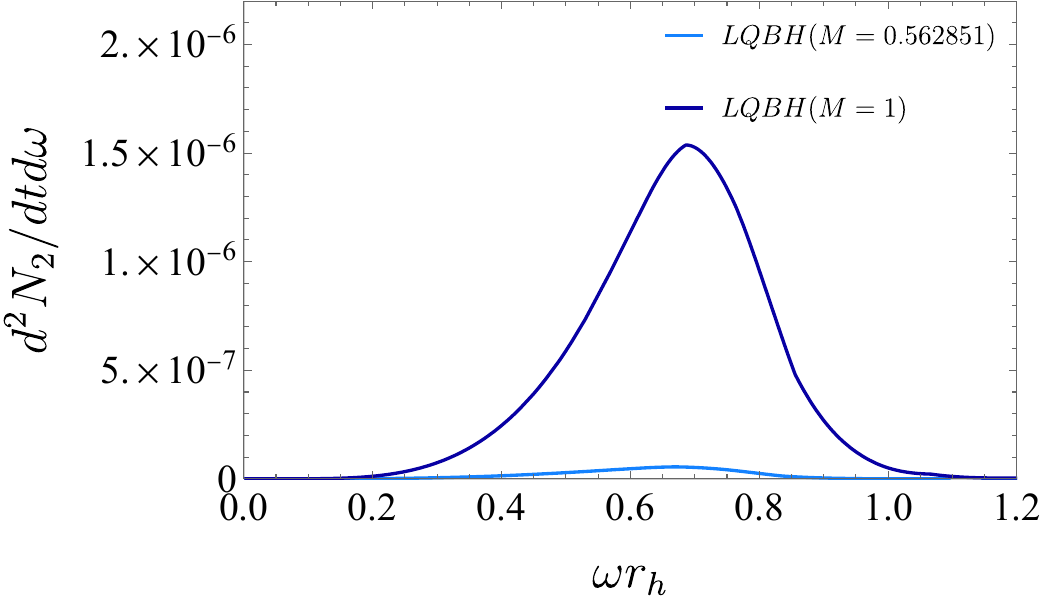}
\end{subfigure}
\caption{The particle emission rates for spins $s=0$, $1/2$, $1$, $2$ as functions of $\omega\, r_h$ for $M=0.56$ and $M=1$ (light blue and dark blue, respectively).}
\label{fig:EmissionRate}
\end{figure}

Once we know the emission of all the particles resulting from Hawking evaporation, we can use the Komar energy ($\ref{komarenergy}$) into the power spectra to obtain a differential equation for the mass loss rate of the black hole,
\begin{equation}\label{evolve}
\frac{dM}{dt}=-\frac{\sqrt{1-r_0/2M}}{{1-r_0/4M}}\frac{dE}{dt}.
\end{equation}
In this work we make use of a method presented in the public code \cite{Blackhawk, Blackhawk2, Blackhawk3} to numerically compute the grey-body factors $\Gamma^{s}_{l}$ for each spin $s$ and mode $l$, and the Hawking evaporation spectra.

Figure \ref{fig:GF} shows grey-body factors for $s=0$, $1/2$, $1$, $2$ with black hole masses corresponding to the maximum temperature and to $M=1$. The mode $l$ increases from $l=s$ going from left to right. We see that it converges to unity as the frequency increases, and the number of $l$ modes in the range $0< \omega r_h<1$ depends on the spin of the field. This means that only lowest modes contribute significantly to the emission rates in the range of re-scaled frequency for the chosen mass, as we can see in Fig.~\ref{fig:EmissionRate}. We find that a decrease in the mass $M$ leads to a decrease in the particle emission rates, and that the emission is more damped for $s=2$.

\begin{figure}[t]
\centering
\includegraphics[width=10cm]{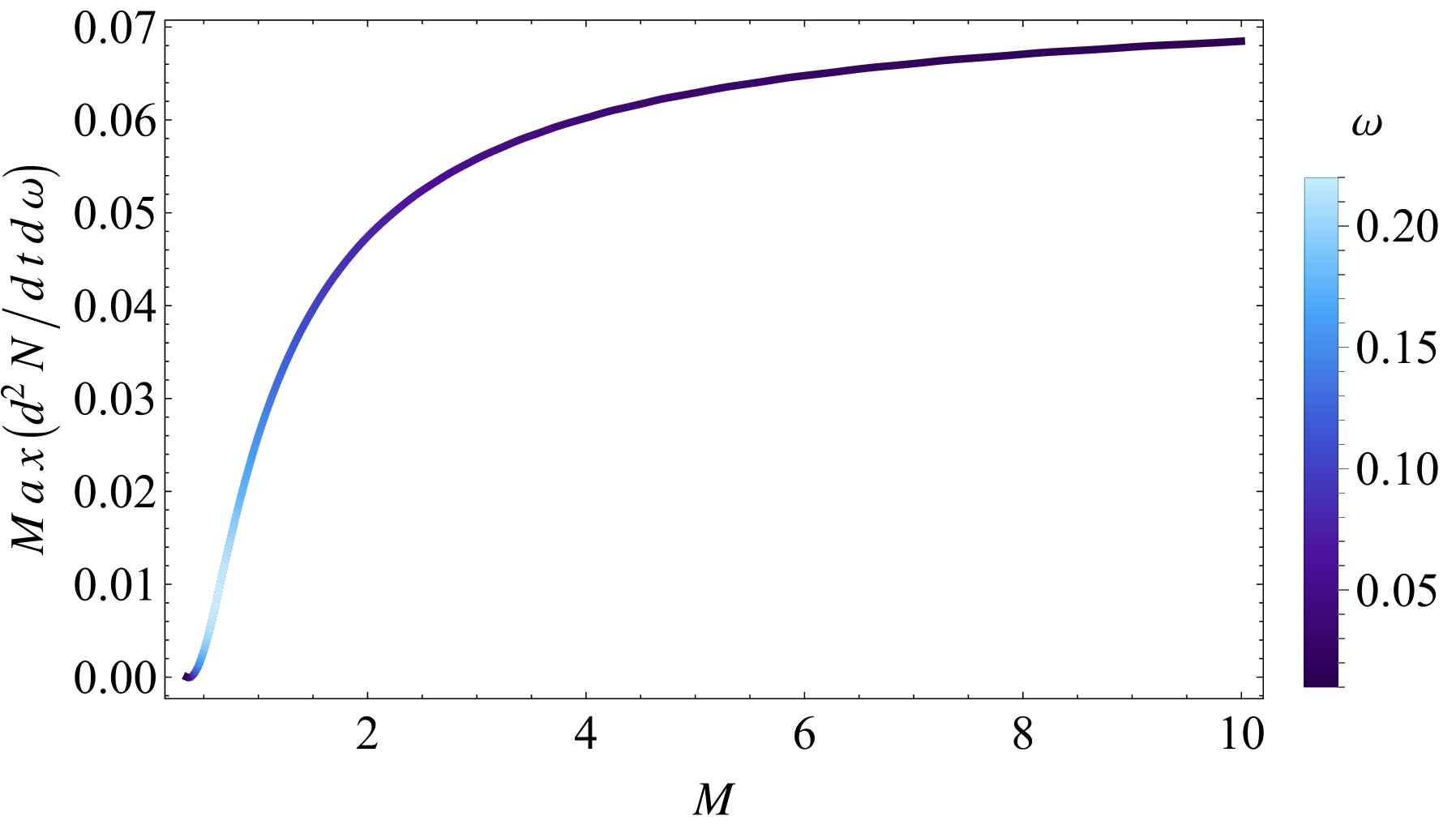}
\caption{\label{fig:maxemissiongradient} The maximum emission rate as function of the black hole mass. The gradient scale shows at which frequency the emission is maximum.}
\end{figure}

Figure \ref{fig:maxemissiongradient} shows the maximum emission rate as a function of the black hole mass. As one can see from the gradient scale, for higher masses the maximum emission rate is centred at very low frequencies. As the black hole emits particles (losing mass), the maximum goes to higher frequencies and, then, goes to zero when the black hole reaches the minimum mass. In other words, the black hole stops the emission at the minimum mass, leaving a remnant.

\begin{figure}[t]
\centering
\includegraphics[width=10cm]{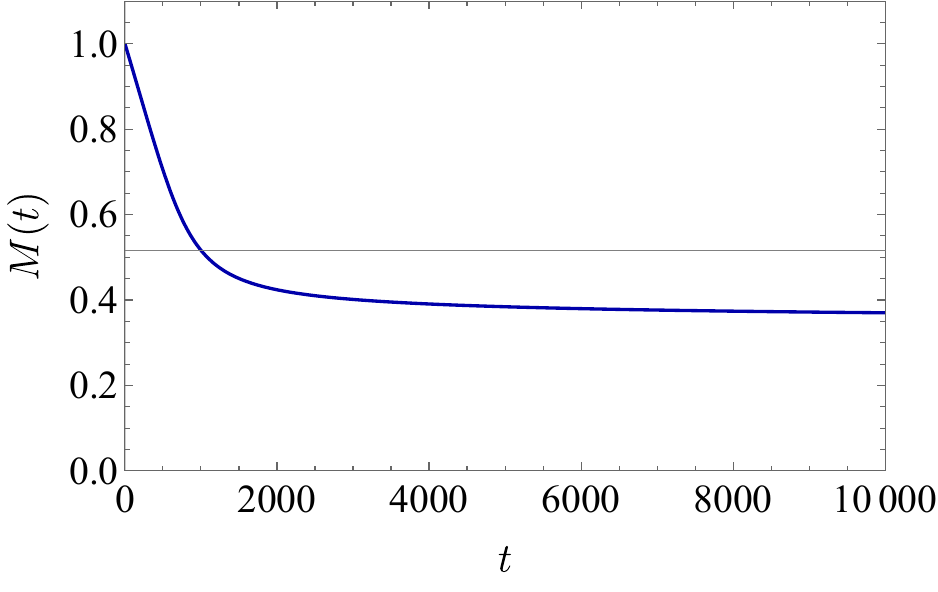}
\caption{\label{fig:Mass_evolution} In blue, the mass evolution for this quantum black hole with initial mass $M=1$. Horizontal grey line represents the mass value where the temperature is maximum.}
\end{figure}

In Fig.~\ref{fig:Mass_evolution} we plot the black hole mass evolution with time, given by expression ($\ref{evolve}$). We can see that the mass of the black hole goes asymptotically to $M_{\text{min}} = r_0/2$, in contrast to the classical solution in which all the black hole mass evaporates.

\section{Concluding remarks}

We have analysed the thermodynamic properties of an effective loop quantum black hole. From the mass dependence of the heat capacity it was shown that the black hole suffers a phase transition when $M=3r_0/4$. In addition,
by considering a scalar field in the vicinity of the horizon, it was possible to obtain the grey-body factors and to calculate the rate of mass loss for this model. The black hole evaporates until reaching the minimum mass $M_{\text{min}} = r_0/2$, in contrast to the classical case, for which a total evaporation is expected. 

There is a fundamental thermodynamic quantity that still deserves to be further investigated, namely the entropy variation of these quantum black holes in their process of evaporation. In a previous study \cite{Fernando2}, the Komar energy and the horizon temperature were used to derive a correction to the Bekenstein-Hawking entropy, assuming an adiabatic evaporation. For a large horizon the classical result is recovered, that is, the entropy equals a quarter of the horizon area. On the other hand, it diverges negatively when the horizon approaches the minimal radius, which was interpreted as balancing the infinite entropy produced in the emission of soft photons when the temperature tends to zero. 

Nevertheless, in the full LQG theory the horizon entropy can be defined as the number of spin network configurations that generate the same horizon area \cite{CQG}. As a horizon with minimal area is pierced by just one spin-network line of colour $1/2$, its entropy is zero according to this counting, which suggests that the final stages of evaporation would in fact constitute a non-adiabatic process.

\section*{Acknowledgements}

The authors are thankful to A. D. Pereira for useful discussions. FGM thanks CAPES (Brazil) for his grant. SC is partially supported by CNPq (Brazil) with grant number 308518/2023-3.


\begin{thebibliography}{}

\bibitem{Wald1999} R. M. Wald, Living Rev. Rel. \textbf{4} (2001) 6.

\bibitem{lewandowski} A. Ashtekar and J. Lewandowski, Class. Quantum Grav. {\bf 21} (2004) R53.

\bibitem{livros1} T. Thiemann, {\it Modern Canonical Quantum General Relativity} (Cambridge University Press, 2008).

\bibitem{livros2} R. Gambini and J. Pullin, {\it A first course in Loop Quantum Gravity} (Oxford University Press, 2011).

\bibitem{livros3} C. Rovelli and F. Vidotto, {\it Covariant Loop Quantum Gravity} (Cambridge University Press, 2015).

\bibitem{PRL} A. Ashtekar, J. Olmedo and P. Singh, Phys. Rev. Lett. {\bf 121} (2018) 241301.

\bibitem{AOS} A. Ashtekar, J. Olmedo and P. Singh, Phys. Rev.
{\bf D98} (2018) 126003.

\bibitem{mariam} M. Bouhmadi-L\'opez {\it et al.}, Phys. Dark Univ. {\bf 30} (2020) 100701.

\bibitem{bojowald} M. Bojowald, Phys. Rev. {\bf D103} (2021) 126025.

\bibitem{espanhois} A. Alonso-Bardaji, D. Brizuela and R. Vera, Phys. Lett. {\bf B829} (2022) 137075.

\bibitem{bascos2} A. Alonso-Bardaji, D. Brizuela and R. Vera, Phys. Rev. {\bf D106} (2022) 024035.

\bibitem{Fernando} F. C. Sobrinho, H. A. Borges, I. P. R. Baranov and S. Carneiro, Class. Quantum Grav. {\bf 40} (2023) 145003.

\bibitem{Fernando2} H. A. Borges, I. P. R. Baranov, F. C. Sobrinho and S. Carneiro, Class. Quant. Grav. {\bf 41} (2024) 05LT01.

\bibitem{bascos3} A. Alonso-Bardaji, D. Brizuela and R. Vera, Phys. Rev. {\bf D107} (2023) 064067.

\bibitem{rakesh} R. Tibrewala, Class. Quantum Grav. {\bf 29} (2012) 235012.

\bibitem{esteban} R. Gambini, E. M. Capurro and J. Pullin, Phys. Rev. {\bf D91} (2015) 084006.

\bibitem{florencia} R. Gambini, F. Ben\'itez and J. Pullin, Universe {\bf 8} (2022) 526.

\bibitem{modesto} L. Modesto, Int. J. Theor. Phys. {\bf 49} (2010) 1649.

\bibitem{Komar1959}
 A. Komar, Phys. Rev. \textbf{113} (1959) 934.

\bibitem{Blackhawk2}
A.~Arbey, J.~Auffinger, M.~Geiller, E.~R.~Livine and F.~Sartini,
Phys. Rev. \textbf{D103} (2021) 104010.

\bibitem{Blackhawk}
A. Arbey and Auffinger, Eur. Phys. J. \textbf{C79}, (2019) 693.

\bibitem{Blackhawk3}
A.~Arbey, J.~Auffinger, M.~Geiller, E.~R.~Livine and F.~Sartini,
Phys. Rev. \textbf{D104} (2021) 084016.

\bibitem{CQG} C. Pigozzo, F. S. Bacelar and S. Carneiro, Class. Quantum Grav. {\bf 38} (2021) 045001.

\end{thebibliography}
\end{document}